\documentclass[12pt,preprint]{aastex}

\shorttitle{Timescale Resolved Spectroscopy of Cyg~X-1}
\shortauthors{Wu et al.}

\begin{document}
\title{Timescale Resolved Spectroscopy of Cyg~X-1}
\author{Y. X. Wu\altaffilmark{1}, T. P. Li\altaffilmark{1,~2,~3}, T. M. Belloni\altaffilmark{4}, T. S. Wang\altaffilmark{2} and H. Liu\altaffilmark{3}}
\altaffiltext{1}{Department of Engineering Physics \& Center for
Astrophysics, Tsinghua University, Beijing, China. E-mail:
wuyx@mails.thu.edu.cn} \altaffiltext{2}{Department of Physics \&
Center for Astrophysics, Tsinghua University, Beijing, China}
\altaffiltext{3}{Particle Astrophysics Lab., Institute of High
Energy Physics, Chinese Academy of Sciences, Beijing, China}
\altaffiltext{4}{INAF-Osservatorio Astronomico di Brera, Via Bianchi
46, I-23807 Merate, Italy}

\begin{abstract}
We propose the timescale-resolved spectroscopy (TRS) as a new method
to combine the timing and spectral study. TRS is based on the time
domain power spectrum and reflects the variable amplitudes of
spectral components on different timescales. We produce the TRS with
the {\it RXTE} PCA data for Cyg~X-1 and studied the spectral
parameters (the power law photon index and the equivalent width of
the iron fluorescent line) as a function of timescale. The results
of TRS and frequency-resolved spectra (FRS) have been compared, and
similarities have been found for the two methods with the identical
motivations. We also discover the correspondences between the
evolution of photon index with timescale and the evolution of the
equivalent width with timescale. The observations can be divided
into three types according to the correspondences and different type
is connected with different spectral state.
\end{abstract}

\keywords{methods: data analysis --- stars: binaries: general ---
stars: individual: Cyg~X-1 --- X-rays: general}

\section{INTRODUCTION}
The X-ray emission from an accreting compact object (neutron star or
black hole) carries information concerning geometry and physical
conditions in the vicinity of the central compact object. One way to
study the X-ray data is to fit various models to the time-averaged
energy spectra. For hard spectral states of black-hole binaries, the
spectra are well understood with a model consisting of weak disk
emission, its Comptonization by a hot corona, and reflection or
reprocessing of the hard X-ray photons by the disk. Other mechanisms
as alternatives to Comptonization, such as jet models, have also
been discussed in the literature \citep[e.g.][]{Mar05,Tom08}. The
spectra of the soft spectral states are characterized with a
dominant soft disk component. In the past few years, efforts have
been made to combine the spectral and variability information to
investigate geometry and dynamics of the X-ray sources. A novel
technique is known as frequency-resolved spectrum or
Fourier-resolved spectrum (FRS), which is based on both power
spectrum and average energy spectrum. This method accumulates the
variability amplitudes (or power spectral density amplitudes) within
a well-defined frequency range for each energy bin to produce the
``energy spectrum''\footnote{The term of ``energy spectrum'' here
might be misleading. Although the FRS has a form similar to the
conventional energy spectrum, they cannot be interpreted in the same
way. The concept will be clarified in the following sections.} for
the specific frequency band. Therefore it provides an opportunity to
explore the variability properties of different spectral components
(e.g. disk emission, power law and iron fluorescent line); as such,
it allows certain immediate insight into the spatial locations or
dynamics responsible for the emission of the specific spectral
components. For example, the FRS can indicate the geometrical size
of the reprocessing medium, because the light crossing time of the
reflector provides a natural frequency filter. Since it was first
proposed by \citet{Rev99}, the FRS has been successfully applied to
Galactic black-hole binaries \citep{Rev99,Rev01,Gil00,Reig06},
neutron star low mass X-ray binaries (LMXBs)
\citep{Gil03,RG06,Shr07} and active galactic nuclei (AGN)
\citep{Pap05,Pap07,Are08}.

In interpreting a Fourier spectrum in the time domain, one usually
takes  $1/f$, the reciprocal of a Fourier frequency $f$, as a
timescale. A time domain power spectrum can be derived directly from
a time series without using the Fourier transform \citep{Li01},
where the definition of power is based only on the original meaning
of rms variation and the power spectrum represents the distribution
of the variability amplitude versus timescale. The Fourier domain
power spectrum is not an accurate representation of rms variations
in the time domain, i.e., for a stochastic process the Fourier
spectrum underestimates the signal power on timescales shorter than
the characteristic time of the process, whereas the time domain
spectrum can correctly estimate it. For the X-ray emission of
black-hole binaries, Fourier spectra and time domain spectra differ
from each other in short timescales or high frequency regions (less
than $\sim0.1$\,s): power densities from time domain spectra are
significantly higher than that from Fourier spectra \citep{LM02}.
For investigating the geometry and dynamics of black-hole binaries,
it is interesting to study the fast variability of the black-hole
binaries X-ray emission by means of timescale-resolved spectroscopy
(TRS) as an alternative to FRS.

In this work, we study the TRS of Cyg~X-1 by using the time domain
power spectrum. The power spectral densities are accumulated within
a certain timescale range in each energy bin, and the ``energy
spectra'' for different timescale ranges are obtained. We introduce
the definition of time domain power spectrum and TRS in Section 2.
The data analysis and results are present in Section 3. In that
section, we first use the same data as analyzed by FRS in earlier
works (Section 3.1), which are in the low-hard (LH) and the
high-soft (HS) state in 1996. In this way we can make a direct
comparison between the two methods, and prove the feasibility of our
new technique. Also in this section we introduce how to check the
data to ensure that the TRS does not suffer systematic biases. In
Section 3.2, we extend the application of TRS to more observations
of Cyg~X-1. We find a correspondence between the evolution of the
power law photon index and the equivalent width of the iron
fluorescent line with timescale. The related issues are discussed in
Section 4.

\section{TIMESCALE-RESOLVED SPECTRUM}
We first introduce the definition of time domain power spectrum. If
$x(k)$ is a counting series obtained from a time history of observed
photons with a time step $\Delta t$, and $r(k)$ is the corresponding
count rate, the variation power is \begin{equation}
\label{eq_pow}P(\Delta t)=\frac{\rm{Var}({\it x})}{(\Delta
t)^2}=\frac{(1/N)\sum_{k=1}^{N}(x(k)-\overline{x})^2}{(\Delta
t)^2}=\frac{1}{N}\sum_{k=1}^{N}(r(k)-\overline{r})^2,\end{equation}
where $\overline{x}$ and $\overline{r}$ are the average count and
count rate, respectively. The power density $p(\Delta t)$ in the
time domain can be defined as the rate of change of $P(\Delta t)$
with respect to the time step $\Delta t$. With two powers at
different timescales, $\Delta t_1$ and $\Delta t_2$ ($\Delta
t_2>\Delta t_1$), we can numerically evaluate the power density at
$\Delta t=(\Delta t_1+\Delta t_2)/2$ by \begin{equation}
\label{eq_pd}p(\Delta t)=\frac{dP(\Delta t)}{d\Delta
t}=\frac{P(\Delta t_1)-P(\Delta t_2)}{\Delta t_2-\Delta
t_1}.\end{equation} For a noise series where $x(k)$ follows the
Poisson distribution, the noise power is
\begin{equation} \label{eq_pn}P_{\rm{noise}}(\Delta t)=\frac{\rm{Var}({\it x})}{(\Delta t)^2}=\frac{\langle
x \rangle}{(\Delta t)^2}=\frac{r}{\Delta t},\end{equation} where
$\langle x \rangle$ is the expectation value of $x$, and $r$ is the
expectation value of the count rate, which can be estimated by the
average count rate of the studied light curve. The noise power
density at $\Delta t=(\Delta t_1+\Delta t_2)/2$ is
\begin{equation} \label{eq_pdn}p_{\rm{noise}}(\Delta t)=\frac{P_{\rm{noise}}(\Delta t_1)-
P_{\rm{noise}}(\Delta t_2)}{\Delta t_2-\Delta t_1}=\frac{r}{\Delta
t_1\Delta t_2}.\end{equation} The signal power density can be
defined as
\begin{equation} \label{eq_pds} p_{\rm{signal}}(\Delta t)=p(\Delta t)-p_{\rm{noise}}(\Delta t),\end{equation}
and the fractional signal power density is \begin{equation}
\label{eq_fpds} p_{\rm{f,signal}}(\Delta
t)=\frac{p_{\rm{signal}}(\Delta t)}{r^2}\end{equation} in the unit
of $\rm{(rms/mean)^2~s^{-1}}$ or s$^{-1}$. In practice we divide the
observation into $M$ segments. For each segment $i$ the fractional
signal power density $p_{\rm{f,signal}}(\Delta t,i)$ is calculated
by equation (\ref{eq_fpds}). Then the average fractional power
density of the studied observation is $\overline{p}(\Delta
t)=\sum_{i=1}^{M}p_{\rm{f,signal}}(\Delta t,i)/M$ and its standard
deviation
$\sigma(\overline{p})=\sqrt{\sum_{i=1}^{M}(p_{\rm{f,signal}}(i)-\overline{p})^2/(M(M-1))}$.

Suppose we have $N_E$ light curves of a given source at different
energy bands $E_j$, for $j=1,2,\ldots,N_E$. For each light curve we
estimate the time domain power spectrum as discussed before. Then
the squared fractional rms can be obtained by integrating the
amplitudes of the fractional power density over the timescale of
interest. For each timescale bin $T_i$, $i=1,2,\ldots,N_T$, TRS can
be constructed according to the formula
\begin{equation} \label{eq_trs} S(T_i,E_j)=C_j\sqrt{[\sum_{t_k\in T_i}p(t_k,E_j)] \delta T},
\end{equation}
where $p(t_k,E_j)$ is the fractional power density at timescale
$t_k$ for the light curve in the energy band $E_j$, $\delta T$ is
the timescale resolution of the time domain power spectrum, $C_j$ is
the count of time-averaged energy spectrum at channel $E_j$,
$S(T_i,E_j)$ represents the amplitude of TRS component in timescale
range $T_i$ and energy band $E_j$. The plot of
$S(T_i,E_j),j=1,2,\ldots,N_E$, as a function of energy constitutes
the ``timescale-resolved spectrum'' or TRS.

\section{DATA ANALYSIS AND RESULTS}
\subsection{The LH and HS States in 1996}
We have applied the TRS as defined above to the publicly available
observations of Cyg~X-1 with the Proportional Counter Array (PCA) on
broad the {\it Rossi X-ray Timing Explorer} ({\it RXTE}). As a first
step, we intended to make a direct comparison between TRS and FRS.
Therefore we chose the same data analyzed through FRS before, which
are observed during the LH state \citep[proposal number
P10238,][]{Rev99} and the HS state \citep[proposal number
P10512,][]{Gil00} in 1996. The specific observation IDs are listed
in Table~\ref{tbl_ob}. Because the single P10512 observations do not
have enough exposure time, we combined the observations that are
close in time (with an abbreviation of 10512 as shown in the first
column of Table~\ref{tbl_ob}). The analysis of single observations
belonging to proposal P10238 displayed similar results. In this
section we take 10238-01-05-000 (abbreviated as 10238b) as an
example.

Because the TRS requires that data have both sufficiently high
energy resolution and good time resolution, we selected the PCA data
in the ``Generic Binned'' mode, which have 16~ms time resolution in
64 energy channels covering 2--100~keV for P10238
(B\_16ms\_64M\_0\_249) and 4~ms time resolution in 8 energy channels
covering 2--13~keV for P10512 (B\_4ms\_8A\_0\_35\_H). We processed
the data with the most recent FTOOLS package (v6.5.1). The data
screening was performed following the criteria that the elevation
angle is larger than $10\,^{\circ}$, the offset pointing is less
than $0.02\,^{\circ}$, all 5 PCU are turned on, and the time since
the peak of SAA passage is larger than 30 minutes. The data were not
filtered on electron ratio as it can exclude valid data for bright
sources. The time-averaged energy spectra for both states were
extracted from the Generic Binned mode and the corresponding
response matrices were created. The full energy resolution
background spectra were extracted from the Standard 2 mode (the PCU
gain correction was applied), and then rebinned into the same
channel configuration as the Generic Binned mode spectra.

We extracted a light curve with a time resolution of 1/64~s in each
energy channel of the Generic Binned mode and produced their time
domain power spectra. Before continuing, we need to do some tests to
guarantee that the time domain power spectra have not suffered
systematic biases, which might be caused by long-term trends,
non-stationarity, background fluctuations and the buffer overflow of
binned data.

\begin{figure}
\plotone{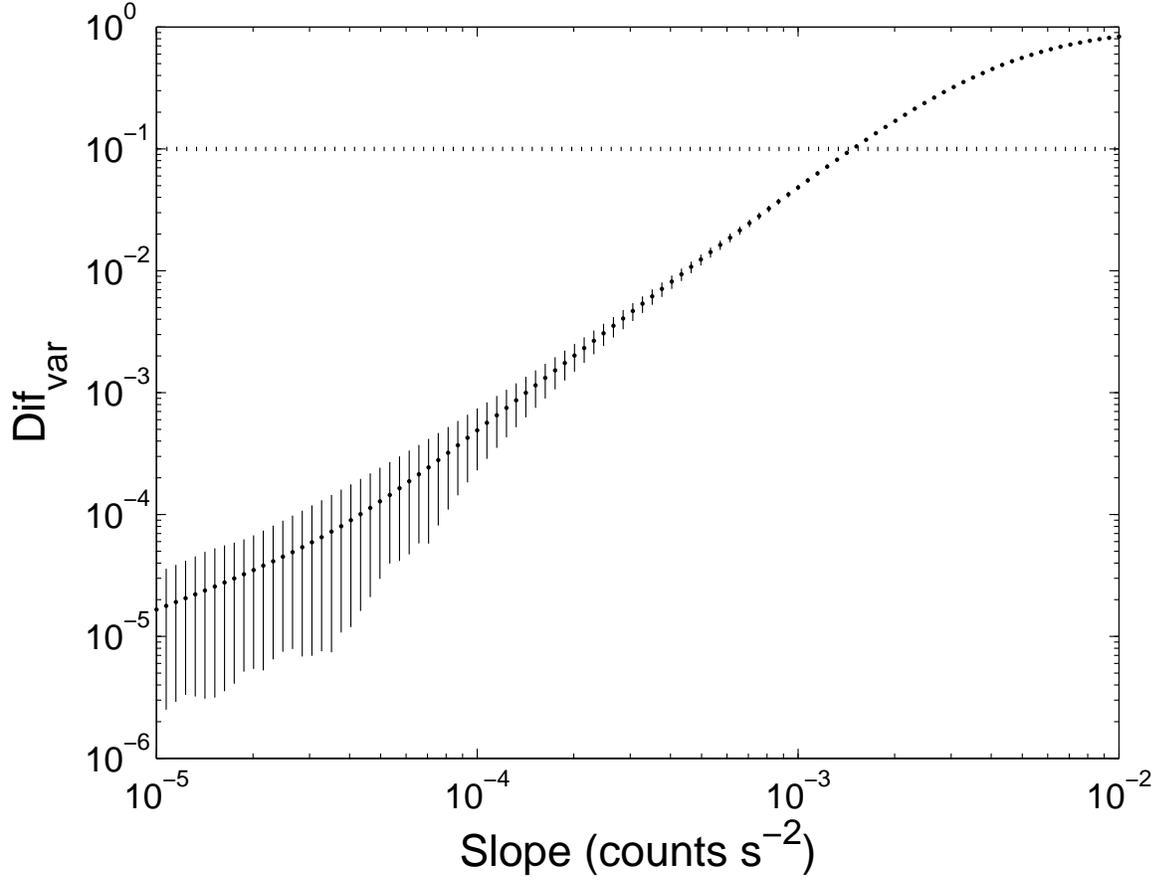} \caption{The relative variance difference
$\rm{Dif_{var}}$ as a function of the slope of the linear trend
derived from simulated light curves. The simulated light curves were
produced by the superposition of a standard normal random series and
a straight line with a certain slope. $\rm{Dif_{var}}$ is the
relative difference between the variance before and after removing
the linear trend. Each point and its error bar is calculated from
100 simulated light curves. \label{fig_sim}}
\end{figure}

The definition of variation power assumes that the time series
fluctuates about a mean count rate, without long-term trends. In our
study we calculated mean variance and power spectra from segments
with a length of 800~seconds, hence the test for the presence of a
trend was performed for each segment. We centered and scaled the
segment light curve by subtracting the mean and dividing by the
standard deviation of the corresponding segment. Then we fitted the
light curve with a linear model. For the observations 10238b and
10512 studied here, we found the best-fitting slope is of the order
of $10^{-4}$ counts~s$^{-2}$. The relative difference between the
variances before and after removing the linear trend is no more than
a few percents, which is significantly smaller than the estimated
error of variation power. More generally, we did a simple simulation
to study the impact of a long-term linear trend to the estimate of
the variance. We produced a light curve with a timestep of 1/64~s by
adding a standard normal distributed random series and a straight
line. We changed the slope of the line and derived the corresponding
difference of variances, as shown in Figure~\ref{fig_sim}. In order
for the relative difference to be smaller than 10\%, the
best-fitting slope should not exceed $\sim 10^{-3}$ counts~s$^{-2}$,
or else the corresponding segment should be excluded or detrended
before the calculation of variation power.

\begin{figure}
\plotone{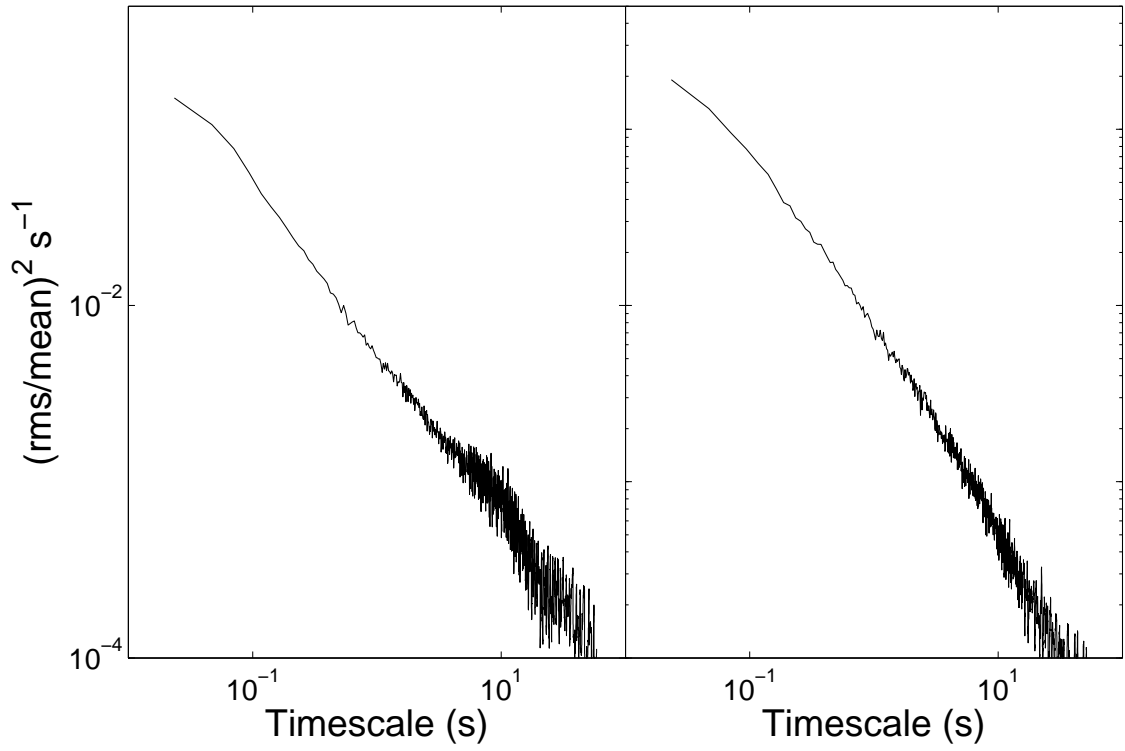} \caption{The time domain power spectra of two
segments of 10512, respectively belonging to the first and last
observation ID. \label{fig_pdsseg}}
\end{figure}

\begin{figure}
\plotone{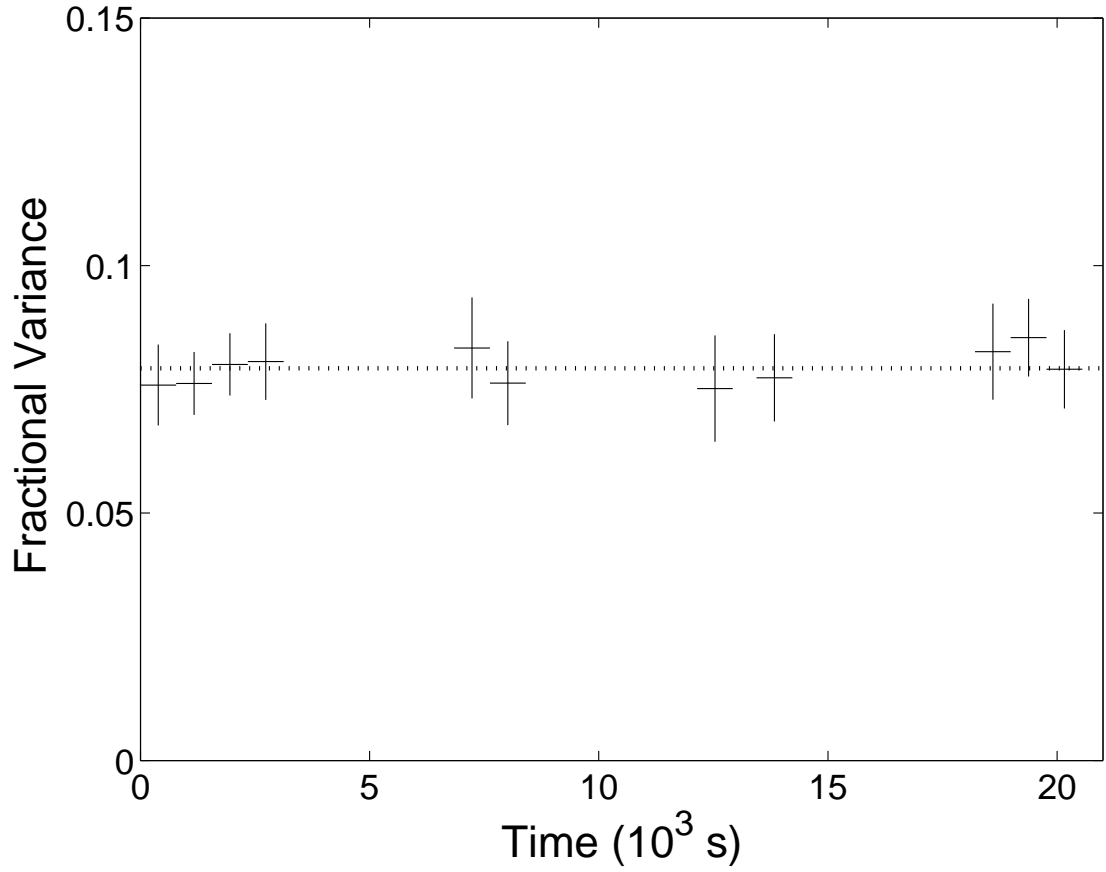} \caption{The fractional variances (variance/flux)
in different segments of 10238b. The dashed line is the mean value.
\label{fig_var}}
\end{figure}

The fractional power density on different timescales is obtained
from each segment, and then they are averaged to reduce random
fluctuations. The average is meaningful only when the intrinsic
statistical properties do not depend on time or, in other words, the
underlying variability processes are stationary so that the scatters
of the power density are simply due to the random fluctuations. The
duration of a pointed observation of {\it RXTE} is usually only
several thousands seconds, but sometimes we have to combine
different observations in order to have enough exposure to produce
the power density. The time separation between observations could be
from hours to days, in which case a test for stationarity is
especially necessary. \citet{Vau03} discussed two approaches for
testing the stationarity of variability, by comparing PSDs and
comparing variances. We performed similar tests to our data. We
derived the power spectra of 10512 from 8 observation IDs of P10512
covering 3 days. The time domain power spectra from the first
segment and the last segment are shown in Figure~\ref{fig_pdsseg}.
No significant difference can be seen. Because it is the fractional
power densities from different segments that are averaged, the
variance divided by the mean flux (fractional variance) from each
segment is compared. Error bars are assigned by measuring the
deviation of multiple estimates \citep{Vau03}. We show the example
of 10238b in Figure~\ref{fig_var}, from which the fractional
variance is consistent with a constant.

The background shows strong fluctuations on large time scales so
that we need to check whether the time domain power spectra are
influenced by it. Cyg~X-1 is a very bright source and, even for the
observation of 10238b in its LH state, the background calibration is
only at a level of $\sim1\%$. We derived the power densities of
background on the timescales of tens of seconds without subtracting
the poisson noise and found they are one order of magnitude smaller
than the source power densities at the same timescale, both for the
entire energy range and for a selected energy range. Furthermore, we
tried subtracting the background light curve from the source light
curve and calculating the time domain power spectrum from the net
light curve. We found it to exhibit a difference generally about a
few percents with the previous one. The role of background
fluctuations is therefore not significant for Cyg~X-1.

\begin{figure}
\plotone{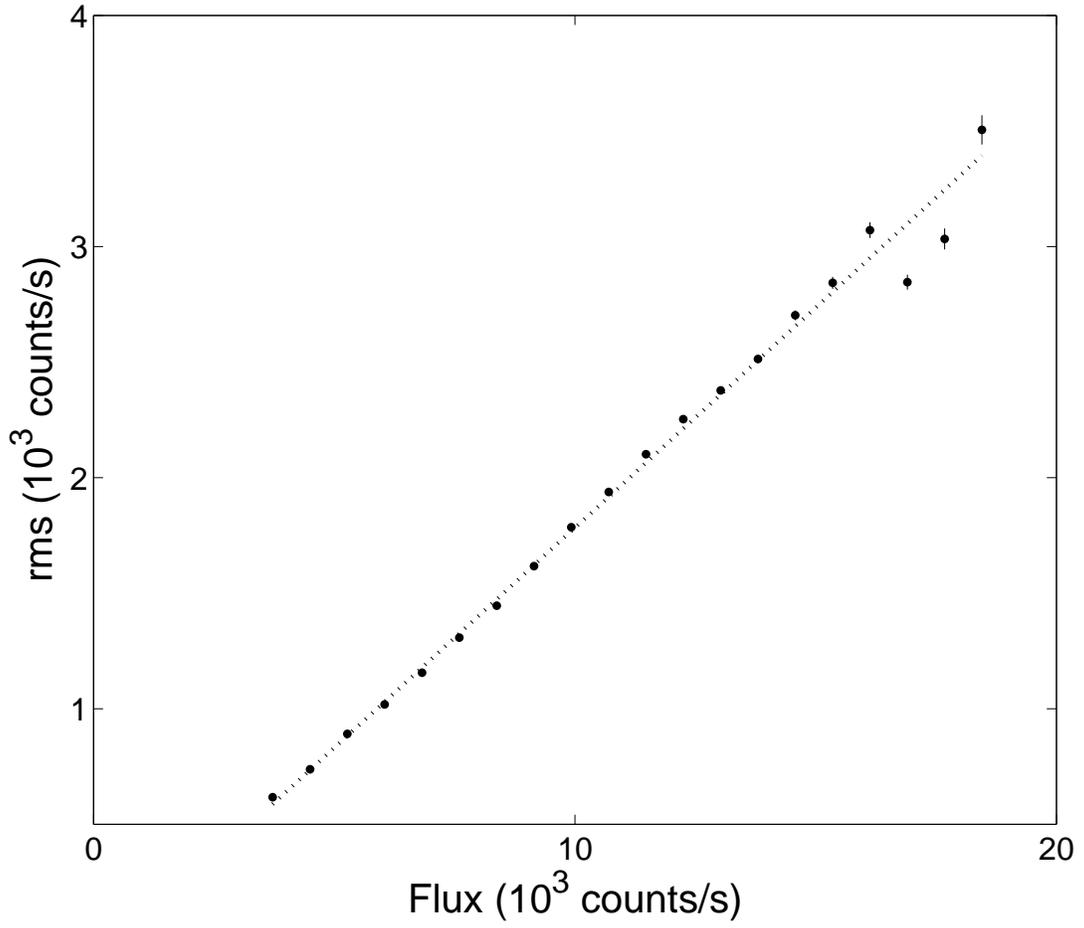} \caption{The rms-flux relation of 10512 in the
channel region of 0--35, corresponding to
2--13~keV.\label{fig_rmsflux}}
\end{figure}

On the other hand, when the source in its HS state at high flux, we
need to take care of a possible buffer overflow suffered by binned
data mode. For the data configuration B\_4ms\_8A\_0\_35\_H, 8-bit
counters are used to form the binned spectrum during each 4~ms
binning interval; that is, up to 256 counts can be accumulated per
4~ms without overflowing. The buffer overflow usually occurs in the
low energy band, in which it will have a remarkable effect on timing
properties such as rms \citep[see e.g., ][]{Gle04}. According to the
{\it RXTE} Technical Appendix, the maximum rate of events that can
be supported without overflowing for the binned mode data used here
is 256,000 counts~s$^{-1}$, which is much larger than the mean count
rate of $\sim8,600$ and the maximum count rate of $\sim42,000$ for
10512. We also derived the rms-flux relation with the method in
\citet{Gle04}, which exhibits a rather good linear relation
(Figure~\ref{fig_rmsflux}). If the data suffered buffer overflow,
the rms-flux relation would display an arch-like shape. The
conclusion is that our data are not effected by buffer overflow.

After checking the data for the possible systematic biases
introduced by the long-term trends, the stationarity, the background
fluctuations and buffer overflow, we ascertained the validity of the
time domain power spectra. Making use of the the time domain power
spectra for each energy channel and background-subtracted energy
spectra, we obtained TRS in 10 timescale bins covering 0.02--60~s
with approximately equal widths in logarithm axis. The uncertainties
of TRS were propagated from the corresponding time domain power
spectra and time-averaged energy spectra.

\begin{figure}
\plotone{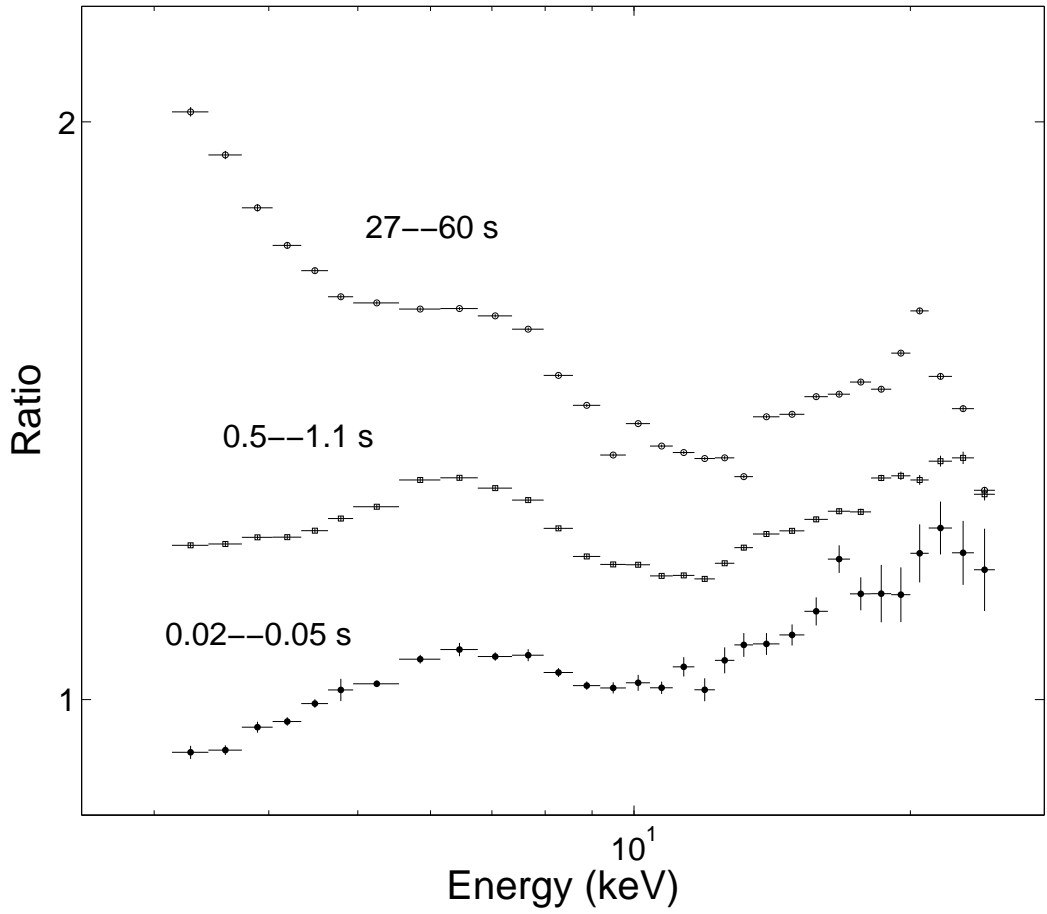} \caption{The ratio of TRS for the timescales of
0.05--0.1~s (dots), 0.5--1.1~s (squares) and 27--60~s (circles) of
10238b to a power law model with the photon index of 1.9. They are
rescaled for clarity.\label{fig_rat}}
\end{figure}

The ratios of TRS to a power law model with photon index of 1.9 are
shown in Figure~\ref{fig_rat}, which visibly exhibits the distinct
characteristics of spectral shapes in different timescale ranges. In
the long timescale range (27--60~s), the TRS has a ``soft excess''
below 4~keV, an broad line around 6--7~keV and a smeared edge above
$\sim 7$~keV. In contrast, the TRS in the short timescale range
(0.05--0.1~s) lacks of the above features below 10~keV, and exhibits
a much more prominent hard component around 20~keV. The shape of the
TRS in the medium timescale range (0.5--1.1~s) is between those in
the long and short timescale ranges, indicating a continuous
evolution of TRS shape with timescale. Therefore it seems that TRS
turns softer and has more prominent feature of iron line, as the
timescale increases.

TRS were fit in the energy range of 3--13~keV with XSPEC v12.3.1. We
selected a simple model consisting of a power law and a Gaussian
line to represent the iron fluorescent line. The low energy
absorption was obtained from the HEASARC website
(http://heasarc.gsfc.nasa.gov/cgi-bin/Tools/w3nh/w3nh.pl) and fixed
at $N_{\rm{H}}=7\times 10^{21}$ cm$^{-2}$. Such a spectral model is
obviously oversimplified, and the reasons we chose it are: 1) the
TRS of 10512 (produced from B\_4ms\_8A\_0\_35\_H configuration) has
only 7 energy channels in 3--13~keV band, restricting the use of
more sophisticated models; 2) the aim of the spectral fit is to
quantitatively measure the observed spectral features, instead of
determining the precise best-fitting model and reveal the true
underlying physics; 3) this is the model used in the previous FRS
analysis for the same data \citep{Rev99,Gil00} and the spectral
parameters can be directly compared with those of FRS. The energy of
iron fluorescent line was fixed at 6.4~keV, and its width was fixed
at 0.6 and 0.8~keV for 10238b and 10512 respectively, which are the
best-fitting values from the corresponding Standard 2 energy
spectra. A uniform systematic error of $0.8\%$ was added to all the
channels when fitting TRS. The free parameters include the
normalization and the photon index of the power law, and the
normalization of the Gaussian line.

\begin{figure}
\plotone{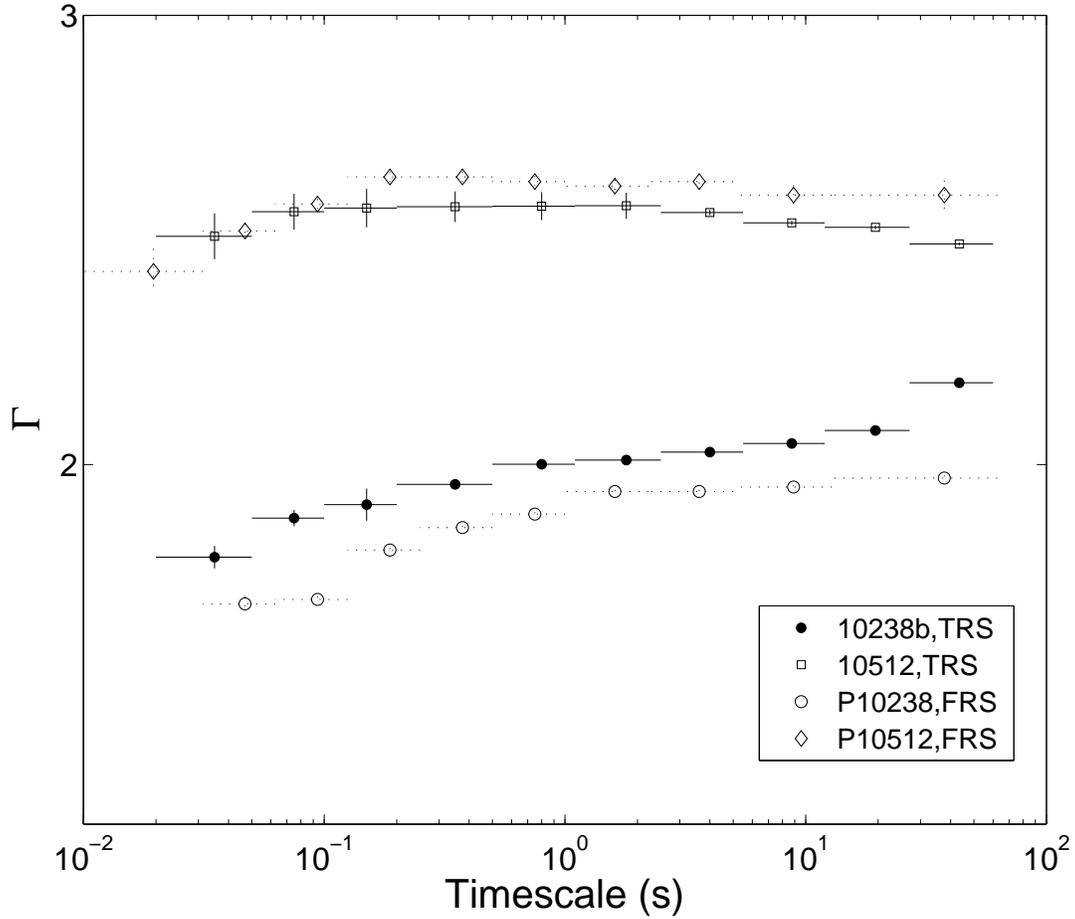} \caption{The dependence of the power law photon
index ($\Gamma$) on timescale for 10238b (dots, solid line) and
10512 (squares, solid line). The results from FRS \citep{Gil00} are
also plotted for the observation P10238 (circles, dotted line) and
P10512 (diamonds, dotted line). \label{fig_idx96}}
\end{figure}

\begin{figure}
\plotone{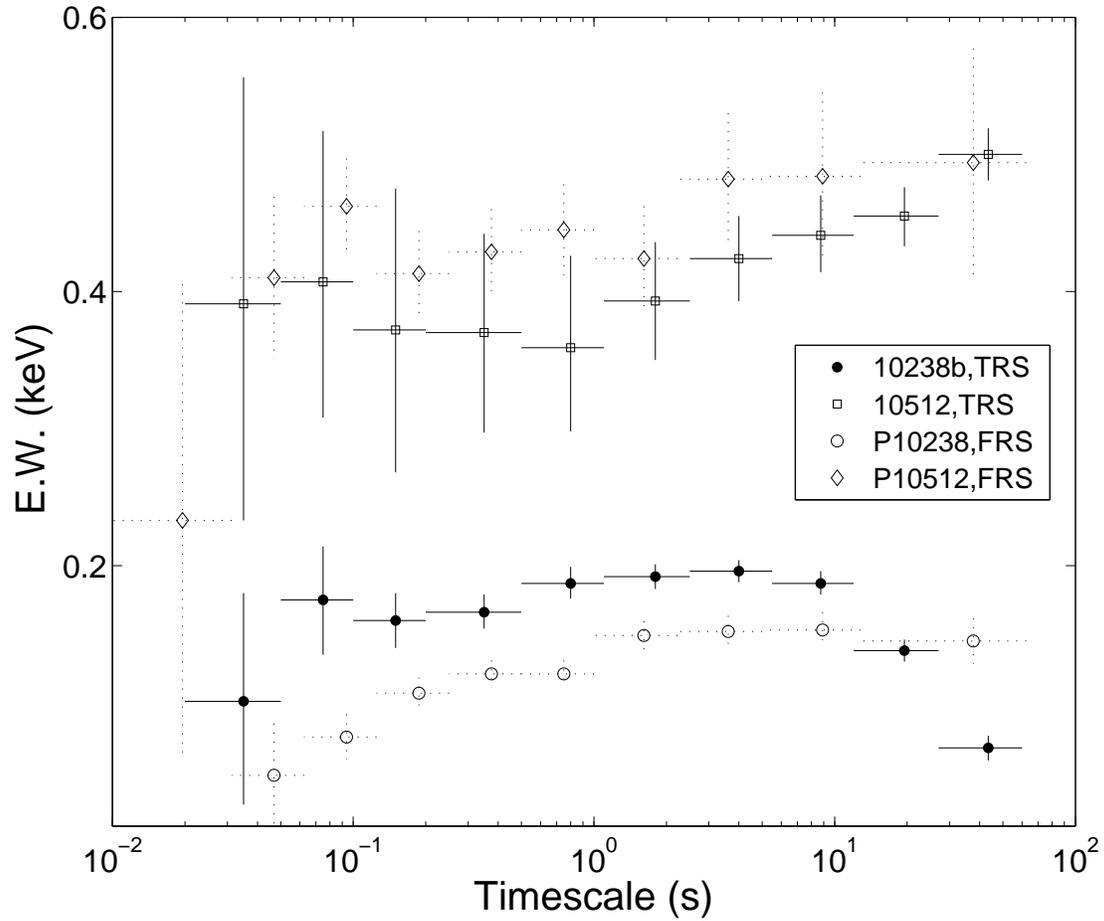} \caption{The dependence of the equivalent width
(E.W.) of the iron fluorescent line on the timescale for 10238b
(dots, solid line) and the 10512 state (squares, solid line). The
results from FRS \citep{Gil00} are also plotted for the observation
P10238 (circles, dotted line) and P10512 (diamonds, dotted line).
\label{fig_eq96}}
\end{figure}

We are interested in the best-fitting power law photon index
($\Gamma$) and the equivalent width (E.W.) of iron fluorescent line,
as listed in Table~\ref{tbl_fit}. The dependence of the above two
parameters on the timescale are plotted in Figure~\ref{fig_idx96}
and Figure~\ref{fig_eq96} respectively. The relation of $\Gamma$ and
timescale exhibits a positive correlation for 10238b. For 10512 it
appears that as the timescale increases, $\Gamma$ increases first,
then keeps stable and finally decreases. However, the relative
amplitude of variation of $\Gamma$ is smaller than 10238b. The E.W.
of 10238b increases with increasing timescale below $\sim$0.1~s, and
stays almost constant in the range 0.1~s--10~s, then decreases above
$\sim$10~s. For the E.W. of 10512, the relative variation amplitudes
are smaller and the uncertainties are larger. The most prominent
difference in comparison with 10238b is that the E.W. does not show
the drop at short timescales, but exhibits a rising trend at long
timescales (several tens of seconds). We also took the data of
$\Gamma$ and E.W. from \citet{Gil00}, translated frequency into
timescale and plotted the result in the same figures. The
similarities are clear. See the discussion section for a specific
comparison between TRS and FRS for the observations around the 1996
state transition.

\subsection{Application to Other Observations of Cyg~X-1}
We have applied TRS to the data of P10238 and P10512, and compared
it with the FRS results from the same data. We discovered that the
two methods do reveal the similar phenomena, which supports the
feasibility of the novel technique of TRS. {\it RXTE} has been
operated for more than a decade and a large amount of observational
data have been accumulated. We tried to extend the application of
TRS to a larger data set.

\begin{figure}
\plotone{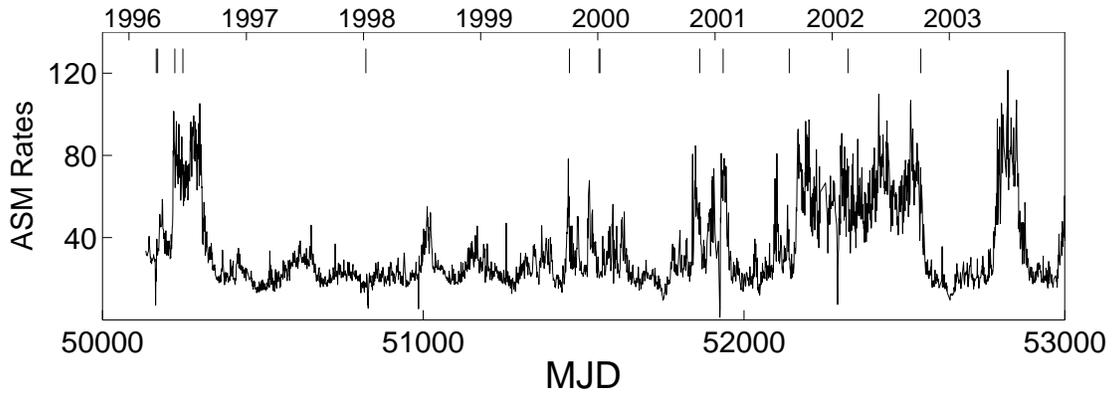} \caption{The ASM light curve covering the
observations in our study. The small, solid vertical lines in the
top indicate the times of observations listed in Table~\ref{tbl_ob}.
\label{fig_asm}}
\end{figure}

To take full advantage of TRS, we require the data with high time
resolution (higher than 1/64~s), good energy resolution (no less
than 8 energy bins covering the channel range 0--35), and enough
exposure time. We searched the {\it RXTE} archive for suitable
observations of Cyg~X-1 and found the Generic Binned configurations
 B\_16ms\_64M\_0\_249, B\_16ms\_46M\_0\_49\_H and
B\_4ms\_8A\_0\_35\_H fulfil our demands. The first two
configurations have only one proposal, which are P10238 and P30157
respectively. The proposals containing the configuration
B\_4ms\_8A\_0\_35\_H are P10214, P10412, P10512, P40101, P40102,
P50109, P60089, P70104. Another very common configuration
B\_2ms\_8B\_0\_35\_Q has only two energy bins between
$\sim$~5--8~keV and is not a best choice for detecting on iron line.
We have produced 11 TRS with the data from the above proposals in
addition to 10238b and 10512 studied in the last section.
Table~\ref{tbl_ob} displays the observation IDs used, as well as the
observation dates and exposure times for each TRS.
Figure~\ref{fig_asm} marks their locations on the overall ASM light
curve. The data analysis followed the same procedures outlined
before.

\begin{figure}
\plotone{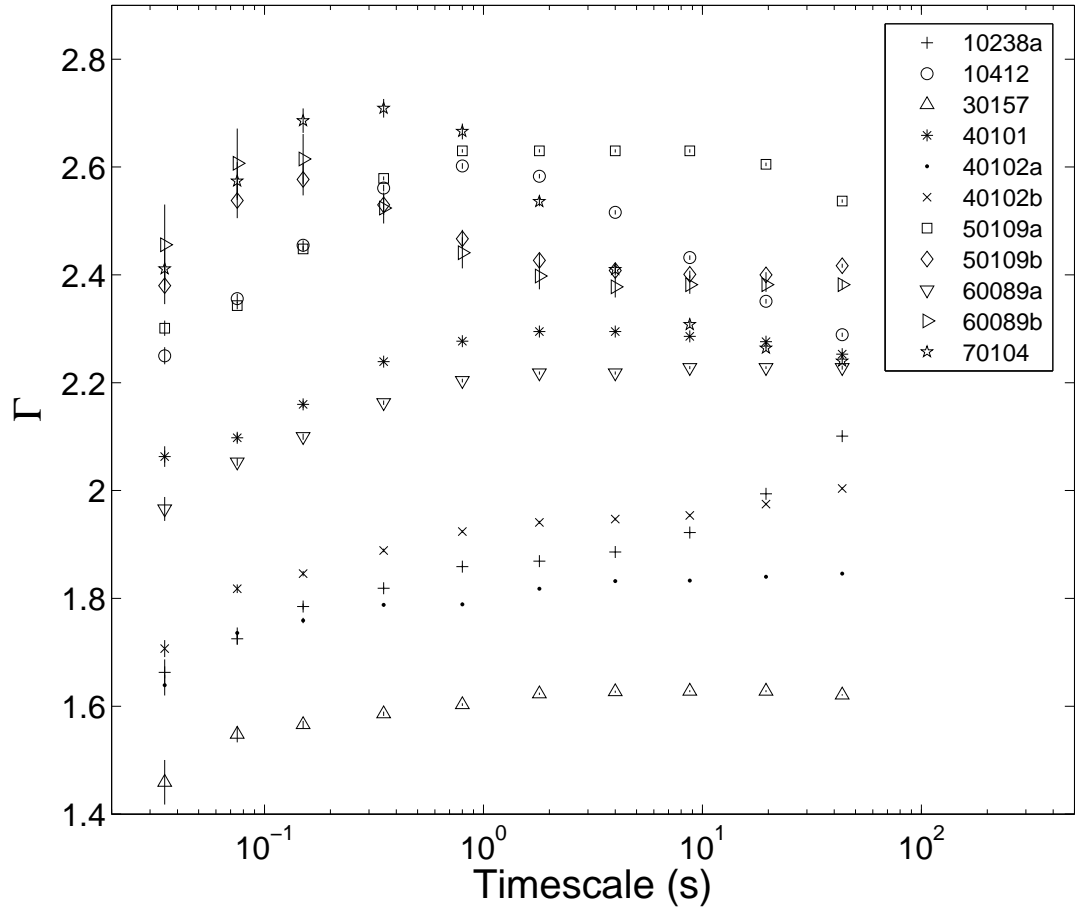} \caption{The best-fitting power law photon index
($\Gamma$) as a function of timescale. \label{fig_idx}}
\end{figure}

\begin{figure}
\plotone{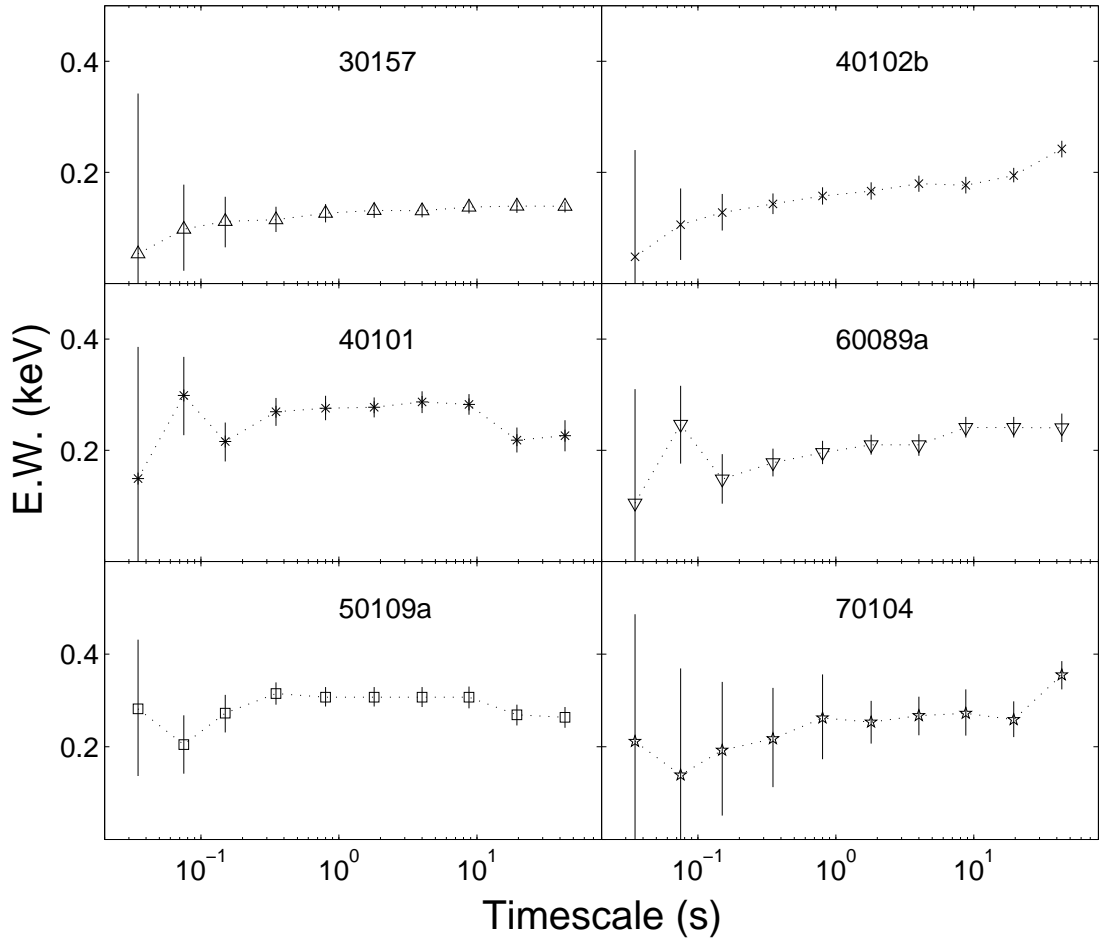} \caption{The equivalent width (E.W.) of the iron
fluorescent line as a function of timescale for six of the
observations. \label{fig_ew}}
\end{figure}

We plotted the power law photon indices for different timescales in
Figure~\ref{fig_idx}. The behaviors of different observations
corresponding to different average $\Gamma$ are different: in the
lower part of Figure~\ref{fig_idx} (or quantitatively for the
observations with maximum $\Gamma < \sim2.2$) $\Gamma$ increases
with timescale until $\sim1$~s, and then remains almost constant
(except 10238a which shows a strong rise at timescales $> \sim10$~s,
similar to 10238b); for the observations with maximum $\Gamma >
\sim2.2$, $\Gamma$ first rises with timescale, then drops down at
larger timescales. This evolution of $\Gamma$ with timescale is
similar to 10512, but the latter is much smoother. The equivalent
widths of iron line on different timescales were also studied and
the representive cases are plotted in Figure~\ref{fig_ew}. The shape
of the relation between E.W. and timescale can be roughly divided
into three types, each corresponding to one row of
Figure~\ref{fig_ew}. The observations with the E.W. evolution shown
in the top row of Figure~\ref{fig_ew} locate in the lower part of
Figure~\ref{fig_idx}. 40101 and 60089a, which have a similar
evolution of E.W. (the middle row in Figure~\ref{fig_ew}), are also
very close to each other in the middle section of
Figure~\ref{fig_idx}. The observations in the upper part of
Figure~\ref{fig_idx}, characterized by the drop of $\Gamma$ at large
timescale, corresponds to the bottom row of Figure~\ref{fig_ew}.
Taking the previous observations, 10238b and 10512 into
consideration, we found 10512 obviously consistent with the
performance of the last type of observations. However, 10238b is
more complex and will be discussed below.

\section{DISCUSSION}
\subsection{Comparison with FRS}
Comparing the results of TRS with FRS for
Cyg~X-1~\citep{Rev99,Gil00}, we found that: 1) the shapes of the
three TRS divided by a power law in Figure~\ref{fig_rat} are similar
to FRS in the corresponding frequency ranges (see Figure~1 in
Revnivtsev et al. 1999 for the ratios of FRS to a power law, in the
frequency ranges of 0.03--0.05~Hz, 4.5--6.8~Hz and 23--32~Hz); 2)
the dependencies of the power law photon index on timescale are
consistent for TRS and FRS, $\Gamma$ increases with the timescale
 for both states when the timescale is below $\sim0.1$~s,
and exhibits a decreasing trend at timescales larger than $\sim10$~s
for the HS state (P10512); 3) the relation of equivalent width of
iron fluorescent line versus timescale obtained from TRS, however,
is inconsistent with that of FRS (see Figure~3 in Gilfanov et al.
2000). In the HS state (P10512), the E.W. drops dramatically at a
timescale of 20--30~ms for FRS, which is absent in the case of TRS.
However, this is probably because TRS does not extend to timescales
as low as FRS (see Figure~\ref{fig_eq96}) and therefore misses the
drop. In the LH state (P10238), the E.W. obtained from both FRS and
TRS drop at timescales shorter than 0.5--1~s. But at longer
timescales, between 10--100~s, the E.W. decreases significantly only
for TRS.

The similar phenomena revealed by TRS and FRS prove that TRS is
another useful tool capable of studying jointly the spectral and
timing properties of X-ray source. It provides the amplitude of
variability in a certain timescale range as a function of energy.
The important practical property of TRS is that they receive the
contributions only from the spectral components that are variable on
the timescale sampled. Therefore by computing TRS we can investigate
whether different spectral components in the average energy spectrum
of the sources (e.g. disk blackbody, power law, iron fluorescent
line) are variable in a given timescale range. Furthermore, the
comparison of TRS with the average energy spectrum could be employed
to study the relative variability amplitudes of various spectral
components at a given timescale.

\subsection{The Variation of 10238b at Large Timescales}
The most remarkable difference between TRS and FRS is that for the
former, the E.W. of 10238b decreases at long timescales ($>$10~s).
What's more, the behaviors of 10238b at long timescales are in fact
unusual compared to other observations. Except 10238a, which belongs
to the same proposal number, we can not find others that display the
drop of E.W. and the rise of $\Gamma$ at timescales of tens of
seconds.

The most straightforward explanation for the variation behavior of
the iron line in the TRS is in terms of a finite light-crossing time
to the distance of the reflector \citep{Rev99,Gil00}. The high
frequency at which the E.W. drops is considered to be the inverse of
the light crossing time between the primary source (assumed as an
isotropic point source above a flat disk in Gilfanov et al. 2000,
and very likely the hard X-ray-emitting corona) and the reflector
(or more specifically, the inner radius of the disk that can process
the X-ray continuum radiation and produce the iron line). With this
characteristic frequency, \citet{Gil00} estimated the inner radius
of the disk to be $R_{\rm{in}} \sim 100~R_{\rm{g}}$ in the LH state
and $R_{\rm{in}} \leq 10~R_{\rm{g}}$ in the HS state. In this
context, the decreases of E.W. at both small and large timescales
may reflect the finite spatial extend of the reflector, for example,
the inner and outer radius of the reprocessing disk. The timescales
at which the equivalent widths drop by a factor of $\geq 2$ are
about 0.04~s and 40~s, corresponding to a distance of $1\times
10^9$~cm and $1\times 10^{12}$~cm respectively. For a 10~$M_{\odot}$
black hole, it might indicate an inner and outer disk radius of
$300~R_{\rm{g}}$ and $3\times 10^5~R_{\rm{g}}$ respectively. The
inner radius is consistent with the transition radius of $\geq
100~R_{\rm{g}}$ in the LH state predicted by Advection-Dominated
Accretion Flow (ADAF) model \citep[e.g.,][]{Esin98}. With a compact
star mass of 10~$M_{\odot}$, a mass ratio of 3 \citep{Gies03} and an
orbital period of 5.6~d \citep{GB82}, we estimated the tidal radius
of accretion disk as $3\times 10^5~R_{\rm{g}}$, also consistent with
the outer disk radius obtained from the dependence of E.W. on the
timescale.

The finite geometry extent of the reprocessing matter is not an
unique way to interpret the variation behaviors of E.W. on different
timescales. Actually it cannot explain why the decrease of the E.W.
at large timescales is not present in other observations,
considering that Cyg~X-1 is a persistent source and therefore the
size of its accretion disk should not have a violent change. The
drop of E.W. at large timescales in 10238b is likely to be caused by
other reasons. Taking notice of the poor goodness of fit for TRS at
the timescale of 27--60~s, as well as the soft spectral shape for
the same timescale in Figure~\ref{fig_rat}, we believe that the
simple model used before is not proper here any more and an
additional soft disk component should be present. The TRS at
27--60~s was fitted again with a model consisting of a disk
blackbody, a power law and a Gaussian line. The reduced $\chi^2$
improved from $331.79/16$ to $27.15/14$, while $\Gamma$ decreased
from 2.18 to 2.09 and the E.W. increased from 0.067~keV to
0.138~keV. It means that after including the disk component in the
model, the striking changes of $\Gamma$ and E.W. at large timescales
disappear. Both parameters stay almost constant at least up to tens
of seconds. Therefore the strange variation of 10238b is probably an
artifact due to the presence of a strong soft component. We suggest
that in this case the disk component is not variable on timescales
shorter than $\sim$10~s, while $\Gamma$ and E.W. keep constant at
the timescale larger than $\sim$10~s. Although the variability in
the disk component was absent in the previous FRS study of Cyg~X-1,
it was indeed found in other sources. In the neutron star LMXB
GX~340+0, the power density spectrum of the disk appears to follow a
power law $P_{\rm{disk}}(f)\propto f^{-1}$, contributing to the
overall variability at low frequency ($<\sim0.5$~Hz) only
\citep{Gil03}. The variation of the disk blackbody component is
related to the viscous instability of the accretion disk and the
variation of the power law spectral component is related to the
thermal instability \citep{Miy94}. The absence of rapid variability
for disk component is consistent with the amplitude of the viscous
timescale \citep[e.g.,][]{Reig06}.

10238a shows similar variations with 10238b at large timescales and
can also be modified by introducing a soft disk component. The other
observations with a much lower energy resolution and degrees of
freedom can not be fitted with a model with an additional disk
component. However $\chi^2$ usually increases at the largest
timescale, which possibly implies the need for a reconsideration of
the applied model. After this modification to the variation at large
timescales, the $\Gamma$ and E.W. of 10238a and 10238b behave alike
other observations.

\subsection{The Classification of Observations}
There exist good correspondences between the relations of $\Gamma$
with timescale and the relation of E.W. with timescale. Based on the
correspondences, we are able to classify the observations analyzed
in this paper into three types.

The first type includes observations 10238a, 10238b, 30157, 40102a,
40102b. The characteristics of this type of observations can be
summarized as (for 10238a and 10238b, the modifications to the
variations at large timescales discussed above have been
considered): 1) $\Gamma$ has a positive correlation with timescale
below $\sim1$~s, and stays almost constant above $\sim1$~s, with the
maximum value of $< 2.1$; 2) E.W. decreases when timescale is
smaller than $\sim0.1$~s, and keeps flat or evolves smoothly at
larger timescale. This type can be readily connected with the LH
state because of the low values of $\Gamma$ \citep[e.g.][]{MR06}.
Indeed all the observations were operated when Cyg~X-1 in its LH
state. 10238a and 10238b were before the state transition occurred
at 1996 May 10 \citep{Cui97,Bel96}, and 30157 was at the long
``quiet'' LH state starting from the end of the HS state in 1996
until 1998 July. Then Cyg X-1 entered a episode with several major
flares. 40102a and 40102b were in the interval of these flares or
``failed state transitions'' \citep[e.g.][]{Pot03}.

The second type contains observations 10412, 10512, 50109a, 50109b,
60089b and 70104. Here: 1) $\Gamma$ increases with timescale below
$\sim0.5$~s, but shows a prominent decrease at timescales above
$\sim1$~s. The maximum values of $\Gamma$ is larger than 2.4; 2) the
E.W. does not decrease significantly at the timescale of
$\sim0.1$~s, instead it may have some increasing trend at short
timescales. This type of observations should be linked with the HS
state according to the large value of $\Gamma$. As to their
observation dates, 10412 and 10512 are observed in the HS state of
1996 (according to the detailed phase division in Cui et al. 1997,
10412 is located at the LH to HS transition phase, but the flux is
at the same level as the following HS state). 50109a and 50109b are
observed during a period of intense flaring, which started in 2000
October and lasted until 2001 March. Both the peak flux and the
minimum hardness ratio reached by the flares are nearly identical to
those of the 1996 state transition, and the source did enter a
period that resembles a sustained HS state during the last outburst
\citep{Cui02}. 60089b and 70104 are observed during the long
2001/2002 HS state staring from 2001 November \citep[e.g.
][]{Zdz02,Wil06}.

The third type includes observations 40101 and 60089a and is
described as: 1) the behavior of $\Gamma$ is similar to the first
type, but with a maximum value around 2.2, and sometimes exhibits a
slight decreasing trend at large timescales, like the second type;
2) E.W. shows a complex behavior at short timescales, increases at
$\sim1$~s but decreases dramatically at smaller timescales. This
type seems like between the first two types, and therefore could be
related with the intermediate state between the LH state and the HS
state. 40101 was observed on 1999 October, when Cyg X-1 was
experiencing a major flare \citep{Pot03}. 60089a is at the initial
stage of 2001/2002 state transition. It is worth to notice that
10238b, although classified into the first type, indeed shows some
similar features with this type, such as the values of $\Gamma$ and
the E.W. at short timescales. 10238b was observed during a minor
flare before the major HS state of 1996, and it is not surprising to
see the similarities because it is not as ``low'' as the other
observations in the first type. So that it appears that the three
types of evolution with timescale have tight connections with the
spectral states, and the translations from one to another are
smooth.

\subsection{Physical Interpretation}
As discussed in the previous section, there exist good
correspondences between the evolutions of $\Gamma$ and E.W. with
timescale, and furthermore they can be connected with the
conventional spectral states of Cyg~X-1. The interpretation for this
is not very clear, and up to now most models involving the X-ray
production mechanism in the accretion process do not consider such
constraints imposed by timescale (frequency) resolved spectral
analysis.

The dependence of E.W. on timescale seems in favor of the finite
geometry size of the reflector as mentioned above, if we focus on
the short timescales. In this picture, the drop of E.W. at
timescales $<\sim0.1$~s in the LH state reflects the finite
light-crossing time between the primary source and the reflector.
However in the HS state E.W. keeps flat down to timescales
$<\sim0.1$~s, which indicates a much smaller distance between
primary source and reflector. This can be explained by the different
inner radius of accretion disk for the two spectral states, just as
stated in \citet{Gil00}.

The spectral slope generally decreases with decreasing timescale or
increasing frequency, which has been proved prevalent in the FRS
study on black-hole binaries, neutron star LMXBs and AGNs. But in
our study it is only true for the timescales shorter than
$\sim0.1$~s to $\sim1$~s depending on the spectral states.
\citet{Pap07} discussed two kinds of models that can qualitatively
explain the fact based on FRS that the spectral shape becomes softer
at larger timescales. One is the phenomenological model attributes
the variable emission to multiple active regions moving towards the
central compact object \citep{Zyc03}. This model assumes that the
Comptonized spectrum evolves from softer to harder during the flare
evolution towards the center with diminishing supply of seed
photons. The other explanation is in the context of ADAF, where the
Comptonization parameter increases with decreasing radius. A
variation at short timescale originates at small radius with large
Comptonization parameter, leading to a hard spectrum. But
\citet{Zyc02} predicted an opposite trend to FRS with the model of
magnetic flare radially distributed above an accretion disk with hot
ionized skin.

\subsection{Application in Weak Sources}
\begin{figure}
\plotone{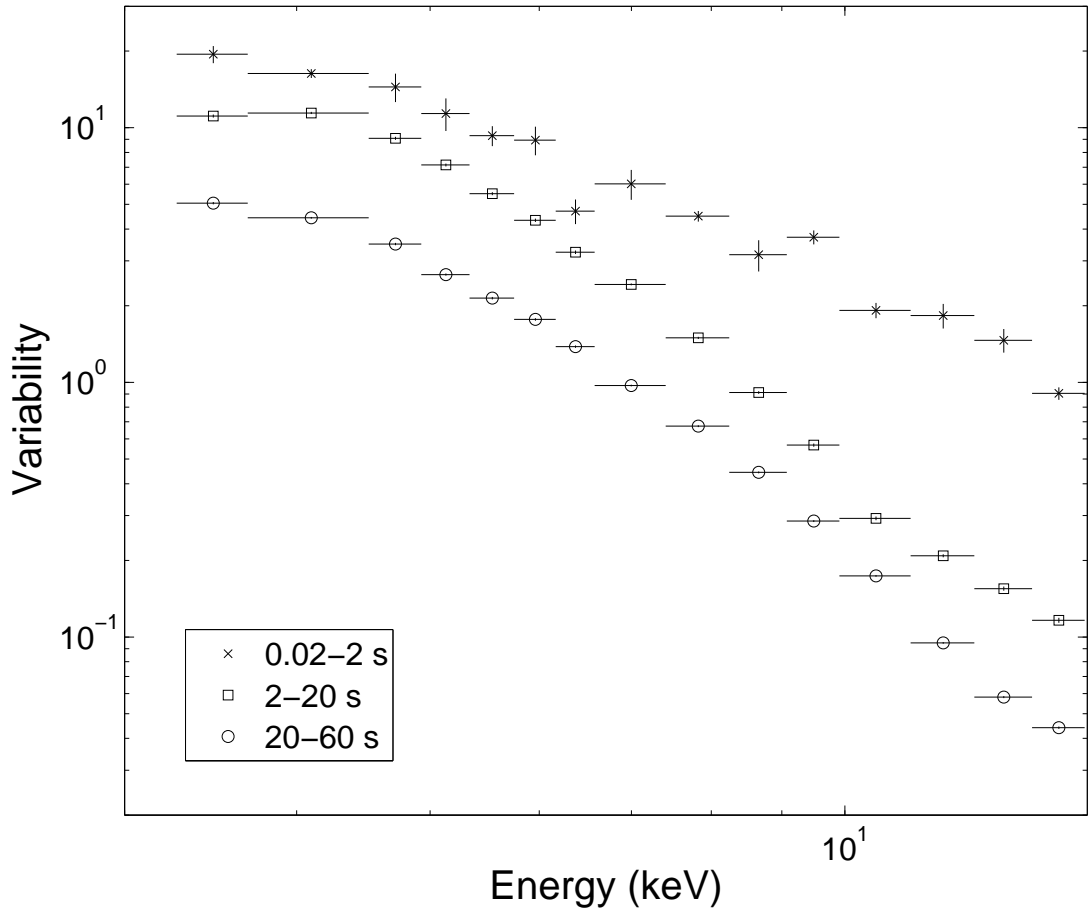} \caption{TRS of 4U~1543-47 in three timescale
ranges. They are rescaled for clarity. \label{fig_4u}}
\end{figure}

Cyg~X-1 is a perfect target for timing studies because of the large
amount of data with different kinds of configurations, its large
flux, and its relatively persistent properties compared with
transient sources. When the method is applied to some weak sources,
there is one additional issue that should be considered. In a faint
source, when power is small, the variability due to the source is
small compared to the variance due to noise, so that subtracting the
expected poisson noise level can sometimes lead to negative
variances due to the fluctuation of the true noise level. This
problem is also met in the study of FRS \citep[e.g.][]{Pap05,Pap07}.
The solution is to sacrifice the frequency (timescale) resolution or
energy resolution to increase the statistical significance. Broad
frequency ranges have to be considered so as to get positive value
after accumulating the power densities in each band, therefore FRS
is usually derived in only two or three frequency bands \citep[see,
e.g.][]{Pap05,Reig06,Pap07,Shr07,Are08}. We show an example of TRS
in 4U~1534-47 (observation ID 70133-01-04-00) in
Figure~\ref{fig_4u}, also studied with FRS in \citet{Reig06}.

In conclusion, TRS is a novel spectral-timing joint analysis
technique, which follows FRS but is based on the time domain power
spectrum. The aims of both TRS and FRS are to determine the
characteristic timescale of the physical process, to probe the
origin of the variability, and to reveal the link between physical
mechanism and geometry arrangement around compact object. Our work
is but a first step. We believe that the application of TRS to more
samples, the study of TRS with higher-quality data, the specific
comparison between TRS and FRS, and the development of theoretical
models will offer a clearer picture and further insight on these
issues.

\acknowledgments  Y. X. Wu thanks the referee for the valuable
suggestions on improving the method, and also wish to acknowledge J.
L. Qu for the help with the data analysis. TMB acknowledge support
from contract PRIN INAF 2006. This work is supported by the National
Natural Science Foundation of China. The data analyzed in this work
are obtained through the HEASARC on-line service provided by the
NASA/GSFC.

\clearpage

\begin{deluxetable}{lllll}
\tablecaption{The list of {\it RXTE} observations used for the
analysis. \label{tbl_ob}} \tablewidth{0pt} \tablehead{
\colhead{Abbreviation} & \colhead{Obs. ID} & \colhead{Start Time} &
\colhead{Exposure (s)} & \colhead{Configuration}} \startdata
\hline 10238a & 10238-01-08-00 & 1996-03-26  10:11 & 23339 & B\_16ms\_64M\_0\_249\\
\hline 10238b & 10238-01-05-000 & 1996-03-30 19:47 & 13381 & B\_16ms\_64M\_0\_249\\
\hline 10412 & 10412-01-01-00 &1996-05-23 14:20 & 7791 & B\_4ms\_8A\_0\_35\_H\\
\hline & 10512-01-07-00 & 1996-06-16 00:01 & 808 &\\
 &10512-01-07-02 & 1996-06-16 04:49 & 1287 &\\
 &10512-01-08-01 & 1996-06-17 01:38 & 686 &\\
 10512&10512-01-08-02 & 1996-06-17 04:50 & 1286 & B\_4ms\_8A\_0\_35\_H \\
&10512-01-08-00 & 1996-06-17 08:02 & 2126 &\\
& 10512-01-09-02 & 1996-06-18 03:15 & 864 &\\
& 10512-01-09-00 & 1996-06-18 06:28 & 1644 &\\
& 10512-01-09-01 & 1996-06-18 09:40 & 2424 &\\
\hline 30157 & 30157-01-05-00 & 1998-01-08 04:37 & 4045 & B\_16ms\_46M\_0\_49\_H\\
& 30157-01-06-00 & 1998-01-15 22:55 & 4364 &\\
\hline 40101 & 40101-01-08-00 & 1999-10-05 07:14 & 3587 & B\_4ms\_8A\_0\_35\_H\\
& 40101-01-09-00 & 1999-10-05 18:33 & 2744 &\\
\hline 40102a & 40102-01-02-00 & 2000-01-05 00:39 & 16386 & B\_4ms\_8A\_0\_35\_H\\
\hline 40102b & 40102-01-01-16 & 2000-01-08 11:48 & 8786 & B\_4ms\_8A\_0\_35\_H\\
\hline 50109a & 50109-03-10-00 & 2000-11-14 06:10 & 9781 & B\_4ms\_8A\_0\_35\_H\\
\hline  & 50109-01-06-00 & 2001-01-26 00:00 & 1557 & \\
50109b & 50109-01-06-01 & 2001-01-26 21:39 & 3145 & B\_4ms\_8A\_0\_35\_H\\
& 50109-01-06-02 & 2001-01-27 20:11 & 1710 \\
\hline 60089a &60089-01-02-00 & 2001-08-20 07:25 & 6084 & B\_4ms\_8A\_0\_35\_H\\
\hline 60089b & 60089-03-07-00 & 2002-02-19 01:08 & 8033 & B\_4ms\_8A\_0\_35\_H\\
& 60089-03-08-00 & 2002-02-19 06:08 & 2254 &\\
\hline & 70104-01-03-00 & 2002-10-03 18:56 & 3561 &\\
70104 & 70104-01-04-00 & 2002-10-04 18:36 & 2427 & B\_4ms\_8A\_0\_35\_H\\
& 70104-01-04-01 & 2002-10-05 17:23 & 2950 &\\
\enddata
\end{deluxetable}
\clearpage

\begin{deluxetable}{cccccccc}
\tablecaption{Best-fitting parameters of 10238b and 10512.
\label{tbl_fit}} \tablewidth{0pt} \setlength{\tabcolsep}{0.04in}
\tablehead{ \colhead{} & \multicolumn{3}{c}{LH
state (10238b)} & \colhead{} & \multicolumn{3}{c}{HS state (10512) } \\
\cline{2-4} \cline{6-8} \\
\colhead{Timescale(s)} & \colhead{$\Gamma$} & \colhead{E.W.(keV)} &
\colhead{$\chi^2$ } & \colhead{} & \colhead{$\Gamma$} &
\colhead{E.W.(keV)} & \colhead{$\chi^2$ } } \startdata
0.02-0.05 & 1.794$\pm$0.025 & 0.101(0.026-0.180) & 9.39 & & 2.508$\pm$0.051 & 0.391(0.233-0.556) & 1.87\\
0.05-0.1 & 1.881$\pm$0.018 & 0.175(0.125-0.214) & 21.14 & & 2.563$\pm$0.040 & 0.407(0.308-0.517) & 1.44\\
0.1-0.2 & 1.911$\pm$0.036 & 0.160(0.140-0.180) & 30.30 & & 2.571$\pm$0.043 & 0.372(0.268-0.475) & 4.33\\
0.2-0.5 & 1.956$\pm$0.009 & 0.166(0.154-0.179) & 23.57 & & 2.574$\pm$0.034 & 0.370(0.297-0.442) & 2.24\\
0.5-1.1 & 2.001$\pm$0.008 & 0.187(0.176-0.199) & 27.44 & & 2.575$\pm$0.031 & 0.359(0.298-0.426) & 2.31\\
1.1-2.5 & 2.010$\pm$0.008 & 0.192(0.183-0.201) & 23.31 & & 2.576$\pm$0.029 & 0.393(0.350-0.436) & 5.78\\
2.5-5.5 & 2.028$\pm$0.008 & 0.196(0.188-0.204) & 20.02 & & 2.561$\pm$0.006 & 0.424(0.393-0.455) & 5.75\\
5.5-12 & 2.047$\pm$0.008 & 0.187(0.179-0.196) & 22.51 & & 2.538$\pm$0.005 & 0.441(0.414-0.470) & 13.58\\
12-27 & 2.076$\pm$0.011 & 0.138(0.130-0.146) & 58.17 & & 2.528$\pm$0.004 & 0.455(0.433-0.476) & 17.54\\
27-60 & 2.182$\pm$0.002 & 0.067(0.058-0.076) & 331.79 & & 2.491$\pm$0.003 & 0.500(0.481-0.519) & 54.49\\
\enddata
\tablecomments{`$\Gamma$'--the power law photon index and its
uncertainty corresponding to 90\% confidence interval; `E.W.'--the
equivalent width of the iron fluorescent line, with its 90\%
confidence interval indicated in the brackets. The degrees of
freedom are 16 and 4 for 10238b and 10512 respectively.}
\end{deluxetable}
\clearpage

\clearpage

\clearpage

\clearpage

\clearpage

\clearpage

\clearpage

\clearpage

\clearpage

\clearpage

\clearpage

\clearpage

\end{document}